\documentstyle[12pt]{article}
\normalsize

\def\a{\alpha}

\def\e{\epsilon}

\def\q{\theta}

\def\t{\tau}

\def\L{\Lambda}

\def\cc{{\cal C}}

\def\cg{{\cal G}}

\def\cz{{\cal Z}}

\def\bar#1{\overline{#1}}

\def\Hat#1{\rlap{\kern.10em$\widehat{\phantom G}$}#1}
\def\HAt#1{\rlap{\kern.05em$\widehat{\phantom G}$}#1}

\def\cap#1{\rlap{\kern.1em$\widehat{\phantom{G\vrule height.8em}}$}#1{}}
\def\Cap#1{\rlap{\kern.05em$\widehat{\phantom{G\vrule height.8em}}$}#1{}}

\let\oldtheequation=\theequation
\def\doteqs#1{\setcounter{equation}{0}
            \def\theequation{{#1}.\oldtheequation}}

\newcounter{sxn}
\def\sx#1{\addtocounter{sxn}{1} \bigskip\medskip \goodbreak \noindent{\large\bf
\centerline{\thesxn.~~#1}} \nobreak \medskip}
\def\sxn#1{\sx{#1} \doteqs{\thesxn}}

\newcounter{axn}

\def\br{\begin{thebibliography}{abc}}
\def\er{\end{thebibliography}}

\date{}

\begin{document}
\bibliographystyle{unsrt}
\footskip 1.0cm
\thispagestyle{empty}
\setcounter{page}{0}
\begin{flushright}
SU-4228-491\\
INFN-NA-IV-91/13\\
UAHEP 9114\\
January 1992
\end{flushright}
\begin{center}{\LARGE THE CHERN-SIMONS SOURCE AS A CONFORMAL FAMILY\\
AND\\
ITS VERTEX OPERATORS\\}
\vspace*{6mm}
{\large A. P. Balachandran,$^{1}$
          G. Bimonte $^{1,2}$ \\
          K. S. Gupta,$^{1}$
          A. Stern $^{2,3}$ \\ }
\newcommand{\bc}{\begin{center}}
\newcommand{\ec}{\end{center}}
\vspace*{5mm}
 1){\it Department of Physics, Syracuse University,\\
Syracuse, NY 13244-1130, USA}.\\
\vspace*{4mm}
 2){\it Dipartimento di Scienze Fisiche dell' Universit\`a di Napoli,\\
    Mostra d'Oltremare pad. 19, 80125 Napoli, Italy}.\\
\vspace*{4mm}
 3){\it Department of Physics, University of Alabama, \\
Tuscaloosa, AL 35487, USA.}\ec

\vspace*{5mm}

\normalsize
\centerline{\bf ABSTRACT}
\vspace*{3mm}

 In a previous work, a straightforward canonical approach to the source-free
quantum Chern-Simons dynamics was developed.  It makes use of neither gauge
conditions nor functional integrals and needs only ideas known from QCD and
quantum gravity.  It gives Witten's conformal edge states in a simple way when
the spatial slice is a disc.  Here we  extend the formalism by including
sources
as well.  The quantum states of a source with a fixed spatial location
are shown to be those of a conformal
family, a result also discovered first by Witten.  The
internal states of a
source are not thus
associated with just a single ray of a Hilbert space.
 Vertex operators for both
abelian and nonabelian sources are constructed.  The regularized abelian Wilson
line is proved to be a vertex operator.  We also argue in favor of a similar
nonabelian result.  The spin-statistics theorem is established for Chern-Simons
dynamics even though the sources are not described by relativistic quantum
fields.  The proof employs geometrical methods which we find are strikingly
transparent and pleasing.  It is based on the research of European physicists
about ``fields localized on cones.''

\newpage
\newcommand{\be}{\begin{equation}}
\newcommand{\ee}{\end{equation}}
\baselineskip=24pt

\sxn{INTRODUCTION}

Theoretical investigations of gauge theories during the past few years have
persuasively established that the Chern-Simons (CS) dynamics is a source of
creative and fertile ideas from a mathematical as well as a physical
perspective.
Its significance for the theory of knots and
links and for an intrinsic
three-dimensional approach to conformal field theories
(CFT's) [1 to 7] is now
widely appreciated.  It is extremely useful for describing fractional
statistics \cite{mor,bal,kum,shan}, while its central
role in theory of quantum Hall effect
(QHE) has also been clarified in recent literature \cite{zhang}.

In  previous communications \cite{bim}, which are reviewed in Section 2,
we investigated the quantization of abelian
and nonabelian source-free CS dynamics by elementary canonical methods
borrowed from
QCD and quantum gravity. When the spatial
slice is a disc $D$, it was shown
that the quantum states form a conformal family localized on the boundary
$\partial D$, a result
discovered first by Witten \cite{witt}.  These are the edge
states of QHE \cite{halp,zhang} when
the gauge group is U(1).  It was suggested in that
paper that our approach has the virtues of great simplicity and transparency as
it does not require scholarship in CFT's or skills in the manipulation of
functional integrals or gauge choices.

In this paper, we extend these considerations by allowing for point sources.
When a point source is immersed in the CS field, it is well known that its
statistics is affected
thereby \cite{mor,bal,kum,shan}. As interaction renormalizes
statistics, it must renormalize spin as well if, as some of us may
conservatively desire, the CS dynamics incorporates the canonical
spin-statistics connection.  The specific mechanism for spin renormalization is
a novel one: the configuration space of particle mechanics is enlarged by a
circle $S^{1}$. A point of $S^1$ can be regarded as parametrizing a tangent
direction or an orthonormal frame (although not canonically). A spinless source
thus ends up acquiring a configuration space which is that of a two-dimensional
rotor with translations added on.  What occurs in CS theory is a conformal
quantum field on this $S^1$ with ability to change its location in space
and with precisely the right spin to maintain the spin-statistics connection.
The necessity for framing the particle has been emphasized before
\cite{witt,shan,tze}. The
qualitative reason for the emergence of this frame
is regularization, which surrounds the
particle with a tiny hole $H$ which
is eventually shrunk to a point.  The CS
action is then no longer for a disc $D$, but
for $D\setminus H$, which is a disc with a
hole.  In contrast to $D$, the latter has an additional boundary $\partial H$,
which is the circle $S^{1}$ mentioned above.
Just as $\partial D$, this boundary as well is associated with a conformal
family.  The internal states of a CS anyon for a fixed location on $D$
thus form an infinite dimensional conformal family of quantum states
and are not described by just a single ray.
This remark was first stated by Witten and applies with equal force to the
quantum Hall quasiparticle \cite{laugh}
if described in the Chern-Simons framework \cite{zhang}.
It is also noteworthy that the CS source is not a first quantized framed
particle, but is better regarded as a ``particle'' with a first quantized
position and a second quantized frame.  One intention of this paper is to
explain these striking, although not quite original results with hopefully
transparent arguments. We undertake this task in Section 2.

The Wilson line integral and its variants have physical and mathematical
importance in gauge theories, the CS dynamics being no exception.  For
instance, the Jones polynomial \cite{jones}
can be understood as the expectation value
of the trace of a Wilson loop with its integration running over a knot or a
link \cite{witt}.  There are also remarks in the
literature \cite{moore,guad,ijmp} roughly to the effect that
the Wilson line emanating on $\partial D$ and terminating at a point $z$ is a
vertex operator \cite{godd} of CFT creating
a source at $z$.
We do not however know
of a published proof showing the correspondence between a Wilson line
and a CFT vertex operator.  We establish it in its abelian form in
Section 3 by
displaying the regularized Wilson line in terms of the Fubini-Veneziano vertex
operator \cite{godd}.

It has been argued elsewhere \cite{alan}
that the canonical spin-statistics connection
does not require field theory or relativity for its validity and holds true in
any dynamical system incorporating creation-annihilation processes subject to
certain rules.  Now the CS dynamics as well is consistent with this connection.
In Section 4, we give a visual and geometric demonstration of this fact using
the properties of the Wilson line.  We will show elsewhere that the CS dynamics
is compatible with the creation-annihilation rules of \cite{alan}
and also present further
developments of these ideas.  Incidentally, arguments
similar to those in this
proof have previously occurred
in work on ``fields localized on cones''\cite{buch}.

Section 5 is our last section and it contains the nonabelian extension of the
preceding work.  In particular, the vertex operators of the nonabelian
Kac-Moody (KM) algebras \cite{godd} are constructed in the CS framework and the
latter are physically interpreted as creation operators of nonabelian anyons
(which are now nonabelian conformal families).  We also establish their
spin-statistics theorem.

There are several publications including our own which quantize CS
sources.  These sources do not get associated with conformal families in many
of
these papers \cite{mor,bal,kum,shan}.  The reason
lies in the rules of quantization
employed in these papers: they are not the same as those which lead to CFT's.
Thus,
if we choose not to identify observables related by nontrivial  gauge
transformations on boundaries such as $\partial D$ or $\partial H$,
we end up with conformal families.  If in contrast, we do
identify these observables, the connection to CFT's is essentially lost, and
the U(1) source for instance ceases to have internal multiplicity of states for
a given location.  Lacking intrinsic reasons for preferring one of these
approaches to quantization and not the
other, we suppose that both will be found to
have their uses to account for appropriate phenomenology.

\newpage

\sxn{CHERN-SIMONS ANYONS AS CONFORMAL FAMILIES}

We have previously remarked that a source of the CS field (as treated here) is
not the same as a framed particle.  For this reason, we refer to the former by
phrases such as a CS anyon or a CS source.

Let us first briefly recall the contents of ref.11, confining ourselves to the
gauge group U(1) and to a spacetime with a disc $D$ as its time slice.
With the convention $\epsilon^{012} = +1$ for the Levi-Civita symbol,
the phase space of the CS action
\be
S=\frac {k}{4\pi} \int_{D \times {\bf R}^{1}} AdA ~~~ A=A_{\mu}dx^{\mu}~~~
AdA=A\wedge dA
\ee
can be described by the equal time Poisson brackets (PB's)

$$\{A_{i}(x),A_j(y)\}=\e_{ij} \frac {2\pi}{k} \delta^{2}(x-y)~;$$
\be
i,j = 1,2~~;~~x^{0}=y^{0}~~;~~\e_{12}=-\e_{21}=1~~;~~\e_{11}=\e_{22}=0
\ee
and the Gauss law constraint
\be
g(\L^{(0)})=\frac {k}{2\pi} \int _{D} \L^{(0)} dA \approx 0
\ee
where
\be
\L^{(0)}|_{\partial D}= 0\;.
\ee
Dirac's weak equality is denoted by the symbol $\approx$ while $\partial D$ is
the boundary of $D$.  The condition (2.4) on the ``test function''
$\Lambda^{(0)}$ is needed if $g(\Lambda^{(0)})$ is to be differentiable in
$A_{i}$ \cite{bim}.  The space of
such test functions will be denoted by ${\cal T}^{
(0)}$. Note
also that $A_{0}$ and its conjugate momentum do not occur in this phase space.
As is permissible, they will hereafter be regarded as nonexistent.

The observables of the theory are
\be
q(\xi)=\frac {k}{2\pi} \int_{D} d\xi A
\ee
(and functions of $q(\xi)$)  where $\xi$ need not vanish on $\partial D$.
Their
PB's at equal times are
\be
\{q(\xi), q(\eta)\}=\frac {k}{2\pi}  \int_{\partial D} \xi d \eta \;.
\ee
(All variables and operators will hereafter be at equal times).  Notice that
test functions $\xi,\xi^\prime$, ... with the same
boundary values define the same observable since
$$q(\xi)- q(\xi^{\prime}) = - g (\xi-\xi^{\prime}).$$

The PB's (2.6) are easily quantized.  Thus let $$\xi_{N}|_{\partial
D}(\theta)=e^{iN\theta},~ N \in {\bf Z},$$
$\theta ({\rm mod}~2\pi)$  being an angular coordinate on $\partial D$
increasing in the anticlockwise sense.
Also, for $D$, we can take $\xi_{0}$ to have the constant value $1$ throughout
$D$. Hence $q(\xi_{0}) \approx 0$ \cite{bim}. We will therefore hereafter
assume that $N \neq 0$ for $D$.
Calling $Q_{N}$ the quantum operator
for $q(\xi_N)$, we find the commutation relation
\be
[Q_{N},Q_{M}] = Nk \delta_{N+M,0}
\ee
which defines a U(1) KM algebra on $\partial D$.  The quantum states are now
constructed by treating $Q_N$ as creation-annihilation operators.  They are
also annihilated by the quantum version ${\cal G}(\Lambda^{(0)})$ of the Gauss
law, ${\cal G}(\Lambda^{(0)})|\cdot >=0$.

Suppose now that a spinless point source with coordinate $z$ is coupled to
$A_{\mu}$ with coupling
$e A_{\mu}(z(x^{0}))\dot{z}^{\mu},~z^{0}=x^{0}$. The field
equation $\partial_{1}A_{2}-\partial_{2}A_{1}=0$ is thereby changed to
\be
\partial_{1}A_{2}-\partial_{2}A_{1}=-\frac {2\pi e}{k} \delta^{2}(x-z).
\ee
If $\cc$ is a contour enclosing $z$ with positive orientation, then, by (2.8),
\be
\oint_{\cc} A =-\frac {2\pi e}{k}.
\ee

On letting $\cc$ shrink to a point, it now follows that
$A(x)=A_{j}(x)dx^{j}$ has
no definite limit when $x$ approaches $z$.
This singularity of $A$ demands regularization.  A good way to regularize is to
punch a hole $H$ containing $z$, and eventually to shrink the hole to a point.

Once this hole is made, the action is no longer for a disc $D$, but for
$D \setminus H$,
a disc with a hole. $D \setminus H$
has a new boundary $\partial H$ and it must be treated
exactly like $\partial D$.  The Gauss law must accordingly be changed to
\be
g(\L^{(1)})\approx 0
\ee
where the new test function
space ${\cal T}^{(1)}$ for $\Lambda^{(1)}$ is defined by
\be
\L^{(1)}|_{\partial D} = \L^{(1)}|_{\partial H}=0 \;.
\ee
The quantum operator ${\cal G}(\Lambda^{(1)})$ for $g(\Lambda^{(1)})$
annihilates all the physical states.

There are now two KM algebras, one each for $\partial D$ and $\partial H$.
The former is defined by observables $q(\xi^{(0)})$ with test functions
$\xi^{(0)}$ which vanish on $\partial H$, the latter by
observables $q(\xi^{(1)})$ with test
functions $\xi^{(1)}$ which vanish on $\partial D$. Let us now define the
KM generators for the outer and inner boundaries as
$$q^{(0)}_{N} \equiv q(\xi_{N}^{(0)})~~,~~ \xi_{N}^{(0)}(\theta)|_{\partial D}
=
e^{iN\theta}~~,~~\xi_{N}^{(0)}|_{\partial H}~=0~;$$
\be
q^{(1)}_{N} \equiv q(\xi_{N}^{(1)})~~,~~~~ \xi_{N}^{(1)}(\theta)|_{\partial H}
=
e^{-iN\theta}~~,~~\xi_{N}^{(1)}|_{\partial D}=0~,
\ee
$\theta$ ( mod 2$\pi$ ) being an angular coordinate on ${\partial H}$.
[ The coordinates $\theta$ on both $\partial D$ and $\partial H$ increase, say,
in the anticlockwise sense. ]
The corresponding quantum operators will be denoted by $Q^{(0)}_{N}$
and $Q^{(1)}_{N}$. Note that the boundary conditions exclude the choice
$\xi_{(0)}^{\alpha} = $ the constant function on $D \setminus H$. Hence we may
not exclude $N=0$ now.

An interpretation of the observables localized on $\partial H$ is as follows.
Let $\theta$ (mod $2\pi$) be an angular coordinate on $D$
which reduces to the $\theta$ coordinates we have fixed on ${\partial D}$ and
${\partial H}$.
A typical $A$ compatible with (2.9) has a blip $-\frac {2\pi
e}{k}\delta (\theta-\theta_{0})d\theta$ localized on $\partial H$
at $\theta_0$.  The behavior
of a general $A$ on $\partial H$ can be duplicated by an appropriate
superposition of these blips.
The observable $q(\xi^{(1)})$ has zero PB with the left side of
(2.9) and hence preserves the flux enclosed by $\cc$.  In fact, the finite
canonical transformation generated by $q(\xi^{(1)})$ changes $A$ to
$A+d\xi^{(1)}$ where the fluctuation $d\xi^{(1)}$ creates zero net flux through
$\cal C$.  All $A$ compatible
with (2.9) can be generated from any one $A$, such as an
$A$ with a blip, by these transformations.  Thus the KM algebra of observables
$Q^{(1)}_{N}$ on $\partial H$ generates
all connections on $\partial H$ with a fixed flux
from any one of these connections.

We have now
reproduced Witten's observation \cite{witt} that the CS anyon or the CS
version of the quantum Hall quasiparticle is a conformal family.

In classical CS theory, there exist diffeomorphism (diffeo) generators which
perform diffeos of boundaries by canonical transformations.  For $\partial D$,
it was shown before \cite{bim} that
they generate the Virasoro algebra and are given by the Sugawara construction
after quantization.  The same is trivially true for $\partial H$.  The $2\pi$
rotation diffeo for $\partial H$,
of interest in the next Section, is readily displayed using
this algebra.

\sxn{THE WILSON LINE IS A VERTEX OPERATOR}

We have seen this statement
occasionally in the literature \cite{moore,guad,ijmp}. Its
correct meaning and proof are likely to be known to some physicists. Here we
demonstrate them in our approach.

The vertex operator \cite{godd} acts
on a state of zero charge (or ``momentum'') and
creates a state with charge.  So let us first establish that the Wilson line
too can be interpreted as playing an analogous role.

For this purpose, take a disc $D$ and pierce a hole $H$ at $z$ as before, but
without inserting a charge at $z$.  There is then a conformal family with zero
charge for this hole, with the ``highest weight state'' or ``vacuum'' $|0>$.
[ The states of this family can be regarded as describing spin fluctuations in
the quantum Hall effect without corresponding charge fluctuations. Cf. Moore
and Read, and Balatsky and Stone [10].]
It is annihilated by the charge operator $Q^{(1)}_{0}$, which is the quantum
version of $q(\xi^{(1)}_{0})$ with $\xi^{(1)}_{0}|_{\partial H} = 1$.  It is
also annihilated by the Gauss law and by $Q^{(1)}_{N}$ for $Nk>0$.
We shall furthermore assume that it
describes the highest weight state of zero charge or vacuum on $\partial D$.

Next consider the Wilson line from a point $P$ on $\partial D$ to $z$, the
integration being along a line $L$:
\be
w(z) = \exp ie \int _{P}^{z}  A\;.
\ee
Its response
$$w(z)\rightarrow [\exp i e \xi^{(1)}_{0}(z)] w(z) \exp [-ie
\xi^{(1)}_{0}(P)]=[\exp i e \xi^{(1)}_{0}(z)] w(z)$$
to the finite transformation $A \rightarrow A + d\xi^{(1)}_{0}$ generated by
$q(\xi^{(1)}_{0})$
shows that it creates a state of charge $e$ at $z$:
\be
Q^{(1)}_{0}w(z)|0>=ew(z)|0>\;.
\ee
(It also creates charge $-e$ at P.)  If the tangent
to $L$ at $z$ points in the angular direction $\theta_{0}$, then
$A_{\theta}d{\theta}$ becomes the blip
$-\frac{2\pi e}{k} \delta(\theta-\theta_{0})d \theta$
at $\partial H$ on the state $w(z)|0>$.  We can say that $w(z)$ creates charge
$e$ localized at $\theta_{0}$ on $\partial H$.

Now $w(z)$ involves a quantum
field on a line and is not a well defined operator.
It requires regularization, a task we can approach as follows.

Let $\Delta$ be a strip of tiny width $\delta$ bounded by $L$ on one side and
by
another line $L^{\prime}$ on the other side (see Figure 1).
[$L$ does not quite reach $z$ here unlike the previous $L$. But this is
immaterial
as $H$ will presently be shrunk to $z$.]
Let $\Lambda$ be
a multivalued function which is
constant on $D \setminus (H \bigcup \Delta)$ and which
increases by $1$ as $\Delta$ is crossed from $L^{\prime}$ to $L$.  Then
$d\Lambda$ is a globally defined closed
one form on $D \setminus H$ and its integral over ${\cal C}$
is $1$. Now consider
\be
\int_{D \setminus H} d \L A.
\ee
It is the same as
$$  \int_{D \setminus (H\cup L)} d \L A \; ,$$
$D\setminus (H \bigcup L)$
being the disc punctured at $H$ and cut also along $L$.  $\Lambda$ is
single valued on this punctured, cut disc and
\be
 \lim_{H \rightarrow z} ~\int_{D\setminus (H\cup L)} d\L A=\int_{L}
A-\lim_{H\rightarrow z} \int_{D\setminus (H\cup L)} \L d A \;.
\ee
In view of the Gauss law, we formally identify (3.3) with
\be
\int_{L}A
\ee
when $H$ shrinks to $z$ and $\delta\rightarrow 0$. [The necessity for the last
limit is not perhaps apparent here.  Without this limit, the exponential of
$ie$ times (3.3) will
create $A$ with support of size
$\delta$ at $\partial H$ and not a blip.]
In this way we are led to consider
\be
W(z)=\exp(ie\tilde{Q}(\Lambda)),~~~~\tilde{Q}(\Lambda) =
\frac{2\pi}{k}Q(\Lambda) = \lim_{H\rightarrow z}\int_{D\setminus H}d\L A
\ee
instead of $w(z)$. [ Here $Q(\Lambda)$ is the quantum operator for
$q(\Lambda)$ ].

Actually let us take one
further step away from $w(z)$, and replace $W(z)$ by
another operator ${\cal W}(z)$ which still creates a blip on $\partial H$,
but creates
uniformly distributed charge instead of a blip on $\partial D$.
For this purpose, let us introduce a function $\Theta$ in the following way.
The angular coordinate $\q$ of a point in $D \setminus H$ is multivalued, but
it can be made single valued by cutting $D \setminus H$ along $L$. The function
$\Theta$
is any single valued determination of $\q$ in $D\setminus (H \bigcup L)$.
On $\partial H$ for example, we can assume that it increases from $\q_{0}$ to
the right of $L$ to $\q_{0} + 2\pi $ to the left of $L$. The operator
${\cal W}(z)$ is obtained from $W(z)$ by putting a new function $\chi$ in place
of $\Lambda$ where i) $\chi$ equals $\Lambda$ on $\partial H$,
ii) $\chi$ becomes $\frac{\Theta}{2 \pi}$ on $\partial D$,
iii) $\chi$ is single valued in $D\setminus (H \bigcup L)$, and
iv) $d\chi$ is a closed globally defined form on
$\underline{D\setminus H}$. [ This means that the integral of $d\chi$
over ${\rm \underline {any}}$
positively oriented closed contour encircling $z$ once is $1$.
If $d \chi$ is replaced by a nonclosed form on $D\setminus H$
(which coincides with
$d \chi$ on $\partial H$ and $\partial D$ ), then the flux
through $\cal C$ for the state
${\cal W}(z)|\cdot >$ will depend on the location and
shape of $\cal C$
and not just its homology class. We can not then regard ${\cal W}(z)$ as
creating charge localized on $\partial H$ and $\partial D$.]
There clearly exists such a $\chi$.

The next step is Fourier analysis.  The function $\chi - \frac {\Theta}{2\pi}$
is single valued on $D \setminus H$
and vanishes as well on $\partial D$.
Only its value
on $\partial H$ is relevant because of the Gauss law.
Its Fourier decomposition on $\partial H$ when $\delta \rightarrow 0$ is
$$\chi (\theta)-\frac {\Theta(\theta)}{2\pi}=
\sum_{N\in\cz}a_{N}e^{-iN\theta}~,$$
\be
a_N=\frac{1}{2 \pi}\int_{\theta_{0}}^{\theta_{0}+2\pi} d\theta
\left (\chi(\theta)
-\frac{\Theta(\theta)}{2\pi}\right )e^{iN\theta}=-\frac{1}{(2\pi)^2}
\int_{\theta_{0}}^{\theta_{0}+2\pi}d\theta \theta e^{iN\theta}.
\ee
Thus
$$a_{0}=-\left (\frac{1}{2}+\frac{\theta_{0}}{2\pi}\right),$$
\be
a_{N}=-\frac{1}{2\pi i} \frac {e^{iN\theta_{0}}}{N}, ~~N \neq 0 \;.
\ee
Hence
\be
\tilde{Q}(\chi)\equiv \int d\chi A = \left [\frac {1}{k}
{Q}(\Theta)-\frac {\pi}{k} Q^{(1)}_{0}\right ]
-\frac{\theta_{0}}{k}Q^{(1)}_{0}+\frac
{i}{k}\sum_{N\neq 0}\frac{1}{N}Q^{(1)}_{N}e^{iN\theta_{0}}\;
\ee
where
\be
{Q}(\Theta) =\frac {k}{2\pi} \int d\Theta A\;,
\ee
and
\be
\left [ \frac{1}{k} {Q}(\Theta)-
\frac{\pi}{k}Q^{(1)}_{0},Q^{(1)}_{N}\right ]=i\delta_{N,0}~~.
\ee

Now $\tilde{Q}(\chi)$ can be written as follows :
\be
\tilde{Q}(\chi)=\frac{1}{\sqrt{|k|}}
\left [q-\epsilon(k)p\theta_{0}+i\sum_{N\neq 0}
\frac{\alpha_N}{N}e^{i\epsilon(k)N\theta_{0}}\right ],~~ \epsilon(k)=
\frac{k}{|k|}.
\ee
Here
$$q=\epsilon(k)
\left [\frac{Q(\Theta)}{\sqrt{|k|}}-\frac{\pi}{\sqrt{|k|}}Q_{0}^{(1)}\right
],$$
$$p=\frac{Q_{0}^{(1)}}{\sqrt{|k|}}~~,$$
\be
\alpha_N=\frac{Q_{\epsilon(k)N}}{\sqrt{|k|}}~~.
\ee
They fulfill the commutation relations
\be
\left [ q,p\right]=i~,~~~\left[\alpha_N,\alpha_M\right]=N\delta_{N+M,0}~~,
\ee
the remaining commutators being zero.
Thus $\sqrt{|k|}\tilde{Q}(\chi)$ is the Fubini-Veneziano coordinate field for
the construction of the vertex operator \cite{godd} with $q$ and $p$ playing
the roles of position and momentum.
Position and momentum, especially the former, occur naturally in CS theory and
need not be introduced by hand.

As in the Fubini-Veneziano construction, we
finally define the correctly regularized vertex operator by the normal ordered
expression
\be
U(\theta_{0})=:\exp~\left(ie\tilde{Q}(\chi)\right ):.
\ee
In this normal ordering, creation and annihilation operators are treated in the
usual way and the $p$
term is moved to the right of the $q$ term.

The Wilson line creates a localized charge on $\partial D$
as well at some angle $\theta^{\prime}_{0}$.  It can also be regularized by the
Fourier analysis of $\Lambda - \frac {\Theta}{2\pi}$ on $\partial H$ and
$\partial D$ and normal ordering.  We find,
$$\tilde {Q}(\L)=\left [ \frac{1}{k}
{Q}(\Theta)-\frac {\pi}{k}Q^{(1)}_{0}\right ]-\frac{\theta_{0}}{k}Q^{(1)}_{0}+
\frac {i}{k}\sum_{N\neq 0}\frac{1}{N}Q^{(1)}_{N}e^{iN\theta_{0}}+$$
\be
-\frac {\pi}{k}Q^{(0)}_{0}-\frac{\theta^{\prime}_{0}}{k}Q^{(0)}_{0}-\frac
{i}{k}\sum_{N\neq 0} \frac{1}{N}Q^{(0)}_{N}e^{-iN\theta^{\prime}_{0}}
\ee
Note that
\be
[Q^{(1)}_{0},{Q}(\Theta)]=-[Q^{(0)}_{0},{Q}(\Theta)]=-i k.
\ee

We can now imagine splitting up
$Q(\Theta)$ as $X^{(0)}+X^{(1)}$, $X^{(0)}$ commuting with
$Q^{(1)}_{N}$ and $X^{(1)}$ with $Q^{(0)}_{N}$.  Then (3.16) becomes the
sum
of two slightly altered Fubini-Veneziano fields with corresponding vertex
operators, the Wilson line becoming the product of these operators.  In the CS
theory, however, the aforementioned split is unnatural as this theory
contains no
operator creating a charge without creating its negative
``image'' charge elsewhere.
\noindent
It is thus
prudent not to insist on the split and write
$${\rm Regularized~ Wilson~ line}~~
=  ~:\exp ie \int^{z}_{P} A:~$$
\be
=~: \exp~\left (ie [\tilde{Q}(\chi) - \frac {\pi}{k}Q^{(0)}_{0}
-\frac {\theta^{\prime}_{0}}{k}Q^{(0)}_{0}-\frac{i}{k}\sum_{N\neq
0}\frac{1}{N}Q^{(0)}_{N} e^{-iN\theta^{\prime}_{0}} ] \right ):~~.
\ee

\sxn{SPIN AND STATISTICS}

As seen in Section 3, the vertex operator (3.18)
which describes the Wilson line creates a pair of
localized charges,
one at the point $z$ in the interior of the disc and another, the ``image''
charge,
at the point $P$ on the outer boundary. In this Section, as a matter of
convenience, the image $P$ for a Wilson line
is taken to be situated at a point $z^{\prime}$ inside
the disc and not on its rim.
In order to regularize the connection $A(x)$ as $x$ approaches $z^{\prime}$,
we create a hole around
$z^{\prime}$ as well and treat it according to the previous Sections.
The images for any two distinct Wilson lines
are taken to be distinct. Note that
the image particles are held fixed and consequently
their internal states are not really important in the
discussion of this Section.

It is useful, for the purpose of analyzing the topic of this Section,
to have a method of
comparing the
internal states of the particle at distinct spatial locations and a
convention which fixes their identity.  We can achieve these goals by
introducing a connection for transporting tangent directions from point to
point.  Let us choose it to be the Levi-Civita connection for a Euclidean
metric on $D$. Directions related by parallel transport will then be said to be
identical.

A two-particle state with identical internal states at locations $z^{(i)}$ is
\be
(:~ \exp ie \int_{L^{(1)}} A~:)~(:~\exp~ ie \int_{L^{(2)}}A~:)~|0>~~,
\ee
$L^{(i)}$ being nonintersecting lines from ${z}^{(i)^\prime}$
to $z^{(i)}$ as in Figure 2.  The tangents to $L^{(i)}$ at
$z^{(i)}$ are parallel, that is, according to the chosen convention,
they point in the same direction.
The vacuum state here is the tensor product of the oscillator vacua for the
holes $H^{(i)}$ and $H^{(i)^{\prime}}$
associated to the two particles and their images.
As $A_{j}$
along $L^{(1)}$ and $L^{(2)}$ commute, we can also write this state as
\be
:~\exp~ ie \int_{L^{(1)}+L^{(2)}}A~:~|0> \;.
\ee

Some qualitative properties of this state will now be pointed out.  If
$L^{(i)}$ are deformed (keeping $z^{(i)},~z^{(i)^\prime}$ and tangents at those
points
fixed), the exponentials are changed by exponentials of the Gauss law and so
the state is not changed.  The lines can even touch as in Figure 3 without
affecting the state.
It must further be observed that the integral over $A$ in (4.2) can be run
along
two lines $L^{(1)^\prime}$ and $L^{(2)^\prime}$ without affecting the state,
the former being from
$z^{(1)^\prime}$ to $z^{(2)}$ and the latter from $z^{(2)^\prime}$
to $z^{(1)}$, as
indicated in Figure 3.  But as these lines intersect, (4.2) can be written in
the form (4.1), with the product of two vertex factors running over
$L^{(i)^\prime}$, only at the cost of an extra phase.

Consider next the adiabatic exchange of the two particles keeping all the
internal states (tangents at $z^{(i)}$ and $z^{(i)^\prime}$) fixed.
The result for the initial state of Figure 4a is
Figure 4b.
It is the same as the successive stages 4c and 4d of Figure 4.
As $L^{(1)}$ and $\bar {L}^{(2)}$ do not have points in common, we can write
the final state as
\be
:~ \exp ie \int_{L^{(1)}}A:~:~\exp ie \int_{{\bar L}^{(2)}}A~:~|0>.
\ee
As the $\bar {L}^{(2)}$ integral is $2\pi$ rotation around $z^{(2)}$
of the $L^{(2)}$ integral,
the spin-statistics theorem has been proved for the state (4.1). Its proof for
a general state can now be constructed along the lines indicated towards the
end of the next Section.

\newpage
\sxn{NONABELIAN GROUPS}

5.1 {\bf The Vertex Operator}

\newcommand{\uG}{\underline G}

We limit our discussion to KM algebras associated with compact simple Lie
groups
$G$. Let $\uG$  be the Lie algebra of $G$.  We identify $\uG$ (or rather
$i\uG$)
with one of its faithful representations $\gamma$  by hermitean matrices.  Let
$\{T_{\alpha}\}$ be a basis for $\gamma$ with
$Tr~T_{\alpha} T_{\beta}=\delta _{\alpha \beta}$.
The connection $A=iA^{\alpha}_{\mu} dx^{\mu}T_{\alpha}$ in our
notation is antihermitian ($A^{\alpha}_{\mu}(x)^{*}=A^{\alpha}_{\mu}(x)$), the
curvature $F=F_{\mu \nu} dx^{\mu} dx^{\nu}$ reads $dA+A^{2}$ and the CS action
is
\be
S=-\frac {k}{4\pi} \int_{D\times R^{1}} {\rm Tr}~ \left ( AdA+\frac
{2}{3}A^{3}\right )\;.
\ee
It leads to the PB's
$$\{A^{\alpha}_{i}(x),A^{\beta}_{j}(y)\}=
\delta_{\alpha\beta}~\epsilon_{ij}~\frac{2\pi}{k}~\delta^{2}(x-y)~~,$$
\be
i,j=1,2~~;~~~x^{0}=y^{0}\;.
\ee

The classical and quantum interaction of nonabelian sources with the CS field
has been treated elsewhere \cite{bal,kum}.  It was
established there that the presence
of a source at $z$ changes the classical Gauss law $F_{12}=0$ to
\be
F_{12}= - i \frac{2\pi}{k}~I~ \delta^{2}(x-z)
\ee
where the hermitean
internal vector $I$ is valued in $\gamma$.  Its range is restricted to an
orbit of $G$ in $\uG$ under the adjoint action, the specific orbit fixing the
unitary irreducible representation (UIR) characterizing the source in quantum
theory.

In the presence of a nonabelian source, the limit of $A$ as $z$
is approached is
uncertain just as  for an abelian source.  This was proved in ref. 3.  It is
not
therefore a matter for surprise that this source term turns into a conformal
family in quantum theory, this being what happens in the U(1) problem.
Briefly,
for quantization, we first pierce a hole $H$ containing $z$ and introduce the
test function space ${\cal T}^{(1)}$ following Section $2$.  Elements $\Lambda
^{(1)}$ of ${\cal T}^{(1)}$ vanish on $\partial D$ and $\partial H$.  They are
now also $\gamma$
valued.  The classical Gauss law is then
\be
g(\L^{(1)})= -\frac{k}{2\pi}\int_{D} {\rm Tr}~ (\L^{(1)}F) \approx 0,
{}~~~~\L^{(1)}\in
{\cal T}^{(1)}\;.
\ee
The quantum analogue ${\cal G}(\L^{(1)})$ of $g(\L^{(1)})$ annihilates all the
physical states.

Observables are generated by two KM algebras, one localized on $\partial D$ and
one on $\partial H$.  They are constructed following Section 2 using test
functions $\xi^{(0)}$ and $\xi^{(1)}$.
(For further details, see ref.11.)  They vanish on $\partial H$
and $\partial D$ respectively and are $\gamma$ valued.  The classical
KM observables
localized on $\partial D$ and $\partial H$ are given by
\be
q(\xi^{(i)})=\frac{k}{2\pi}\int_{D}{\rm
Tr}~(-d\xi^{(i)}A+\xi^{(i)}A^{2}),~~~~i=0,1.
\ee

We now focus attention on $\partial H,~\partial D$ having been looked at
before \cite{bim}.
The quantum
KM generators $Q^{(1)\alpha}_{N}$ localized on $\partial H$ are the
quantum operators for
 $$q(\xi^{(1)\alpha}_{N}),~~~~\xi^{(1)\alpha}_{N}|_{\partial
H}(\theta)=ie^{-iN\theta}T_{\alpha}~~({\rm and}
{}~\xi^{(1)\alpha}_{N}|_{\partial D}=0).$$  They fulfill the algebraic relation
\be
[Q^{(1)\alpha}_{N},Q^{(1)\beta}_{M}]=i
f^{\alpha\beta\gamma}Q^{(1)\gamma}_{N+M}+Nk\delta_{N+M,0}~\delta_{\alpha\beta}
\ee
and the hermiticity
property $Q^{(1)\alpha^{\dagger}}_{N}=Q^{(1)\alpha}_{-N}$ in a unitary KM
group representation.

Given an algebra such as (5.6), there is a standard approach to finding its
irreducible representations \cite{godd}.  Let $H_{i}(1\leq i \leq r),
E_{\alpha}$ be the Cartan
or rather the Chevalley  basis for $\uG$ with $H_{i}$ spanning the Cartan
subalgebra and $\alpha >0$ denoting the positive roots.  Let
$H^{(1)i}_{N},~E^{(1)\alpha}_{N}$ denote the quantum operators
$Q^{(1)}(ie^{-iN\theta}H_{i})$, $Q^{(1)}(ie^{-iN\theta}E_{\alpha})$ for
$q(ie^{-iN\theta}H_{i})$, $q(ie^{-iN\theta}E_{\alpha})$
where in the arguments, only the
boundary values of test functions $\xi^{(1)}$
on $\partial H$ have been displayed for simplicity.  Then
$H^{(1)i}_{N},~E^{(1)\alpha}_{N} (N>0)$ and $E^{(1)\alpha}_{0} (\alpha >0)$ are
the generators associated with the
positive roots of the (extended) KM algebra whereas $H^{(1)i}_{0}$ span the
Cartan subalgebra.  A representation is obtained by setting up the highest
weight
state $|h>,~h=(h^{1},h^{2},...,h^{r})$, fulfilling $H^{(1)i}_{0}|h>=h^{i}|h>$
 and
annihilated by all the positive root generators ~:
$H^{(1)i}_{N}|h>=E^{(1)\alpha}_{N}|h>=0
$ if $N>0$ and $E^{(1)\alpha}_{0}|h>=0$ if $\alpha>0$. $h$~ is
also subject to certain
constraints \cite{godd}. Repeated applications
of the negative  roots then generate
all the states of the representation.  There is a scalar product guaranteeing
the hermiticity conditions
$H^{(1)i^{\dagger}}_{N}=H^{(1)i}_{-N},~E^{(1)\alpha^{\dagger}}_N=E^{(1)-\alpha}
_{-N}$.

In our case, there is a conformal family sitting on $\partial D$ even prior to
the insertion of a source.  We will assume that $|h>$ is also a highest weight
state (say) for $\partial D$ and will not bother to display this fact.

We want to describe the insertion of a source using a vertex operator as in the
U(1) CS theory.  It is known that a vertex operator exists only for level
$k = 1$
KM representations \cite{godd}.  Let us specialize to $G=SU(2)$ for simplicity.
Then $r=1$ and $h^{1}$ (the conventional ``third component of angular
momentum'')  for a highest weight state
is $0$ or $\frac{1}{2}$ for a level 1
representation.( Here and in what follows
until (5.19),
we have taken the longest root of $SU(2)$ to be
normalized to 1.) These two representations are inequivalent. The basic vertex
operator we construct generates the
$h^1 = \frac{1}{2}$ highest weight state from
the $h^1 = 0$ highest weight state. It is associated with the creation of a
particular sort of source. It performs this creation in a way similar to that
of an abelian vertex operator of Section 3. Other related vertex operators we
discuss have analogous properties.

It is convenient to identify $SU(2)$ with $2\times 2$ unitary unimodular
matrices.  Then we can set
$$H_{1} := \frac{1}{2}\tau_{3},~~~E_{\a}(\a>0) :=\t_{+}=\frac {1}{2}(\t_
{1}+i\tau_{2}),~~~E_{\a}(\a<0) :=\t_{-}=\frac{1}{2}(\t_{1}-i\t_{2}),$$

$$H^{(1)3}_{N}:~=Q^{(1)}(ie^{-iN\q}\frac{\t_{3}}{2}),$$
$$E^{(1)\a}_{N}(\a>0):~=E^{(1)+}_{N}=Q^{(1)}(ie^{-iN\q}\t_{+}),$$
\be
E^{(1)\a}_{N}(\a<0):~=E^{(1)-}_{N}=Q^{(1)}(ie^{-iN\q}\t_{-})\;.
\ee
$H^{(1)1}_{N}$ has been renamed here as $H^{(1)3}_{N}$ and $\tau_{i}$ are Pauli
matrices.

We can now go about
finding a vertex operator method of source insertion as in
the abelian CS theory, partially
imitating considerations in KM theory.  A hole $H$
is first made at $z$ and a highest weight state $|0>$ with eigenvalue $0$ for
$H^{(1)3}_{0}$ is first erected on $\partial H$.  It is annihilated by the
generators for the positive roots and by $H^{(1)3}_{0}$.

Now from a KM highest weight state, we can uniquely induce a unitary
irreducible representation (UIR) of SU(2). For it is annihilated by its raising
operator $E^{(1)+}_{0}$ and is its highest weight state as well.  As $E^{(1)+}_
{0}$ and $H^{(1)3}_{0}$ kill $|0>$, this UIR is the trivial one for $|0>$, and
$E^{(1)-}_{0}$ as well annihilates it.

Let the internal vector $I$ of the source determine the ``total angular
momentum'' $j=\frac{1}{2}$ UIR of SU(2). The KM highest weight state $|\frac
{1}{2}>$ is clearly associated with this SU(2) UIR.  We shall later discuss the
sense in which (5.3) is enforced on the
two states $|\frac{1}{2}>$,
$E^{(1)-}_{0}|\frac{1}{2}>$ with $j=\frac{1}{2}$ and $H^{(1)3}_0 =
\pm\frac{1}{2}$
and also how to create $j$ which differs from $\frac
{1}{2}$.  We are now ready to reveal the operator $V$ which creates $|\frac
{1}{2}>$ from $|0>$.

Let $\underline{\theta} =
i \theta \frac{\tau_3}{2}$ where the value of the function
$\theta$ at a point is the polar coordinate at that point. This function
is well defined on $D\setminus (H \bigcup L_0)$
where the line $L_0$ from $\partial D$ to $\partial H$ has zero polar angle.
Consider
\be
Q(\underline{\theta}) =
\frac {1}{2\pi}\int {\rm Tr}~(-d\underline{\theta} A+\underline{\theta} A^{2})
\ee
where $k$ now is being set equal to $1$. Its commutator with $Q(\xi^{(1)})$ is
\be
[Q(\underline{\theta}), Q(\xi^{(1)})]=
-iQ \left ( [ \underline{\theta}, \xi^{(1)}]\right )
-\frac{i}{2\pi}\int_{\partial H}{\rm Tr}~\xi^{(1)}d\underline{\theta}~~.
\ee
Special consequences of this equation are
\be
[Q(\underline{\theta}),H^{(1)3}_{N}]=\frac{i}{2}\delta_{N,0}~~,
\ee
\be
[Q(\underline{\theta}),E^{(1)\pm}_{N}] = Q\left(\pm\theta i
e^{-iN\theta}\tau_{\pm}\right )~.
\ee

{}From $Q(\underline{\theta})$, we form the operator
\be
V = e^{iQ(\underline{\theta})}~.
\ee
It creates a state $V|0>$ of uniform $\frac{\tau_3}{2}$ charge $\frac{1}{2}$ on
$\partial H$ ( and $-\frac{1}{2}$ on $\partial D$) since by (5.10),
\be
H^{(1)3}_{0}V|0>=\frac{1}{2}V|0> ~~.
\ee
Furthermore, by (5.10) and (5.11),
$$V^{-1}H^{(1)3}_{N}V=H^{(1)3}_{N},~~~N\neq 0~~,$$
\be
V^{-1}E^{(1)\pm}_{N}V=E^{(1)\pm}_{N\pm 1}~.
\ee
Hence,
$$H^{(1)3}_{0}V|0>=E^{(1)\pm}_{N}V|0>=0~,~~~N>0~,$$
\be
E^{(1)+}_{0}V|0>=0~.
\ee
$V|0>$ is thus the highest weight state $| \frac {1}{2}>$
of the entire KM algebra
and induces the $j=\frac {1}{2}$ UIR of SU(2).

The state $|\frac {1}{2}>$ is gauge invariant.  (Hence so are all the level 1
KM states.)  To see this, let us regard $\Lambda^{(1)} \in {\cal T}^{(1)}$ in
the Gauss law
${\cal G}(\Lambda^{(1)})$ as
valued in the SU(2) spin $\frac{1}{2}$ Lie algebra and
write $\Lambda^{(1)}=\Lambda^{(1)}_{-}
\tau_{+}+\Lambda^{(1)}_{+}\tau_{-}+i\Lambda^{(1)}_{3}\frac {\tau_{3}}{2}$.  We
have
\be
V^{-1}\cg (\L^{(1)})V=\cg
(\bar{\Lambda}^{(1)})~,~~~\bar{\Lambda}^{(1)}=\L^{(1)}
_{-}e^{-i\q}\t_{+}+\L^{(1)}_{+}e^{i\q}\t_{-}+i\L^{(1)}_{3}\frac
{\tau_{3}}{2}\;.
\ee
The function $\bar {\Lambda}^{(1)}$ is well defined on $D\setminus H$ (being
the same for $\theta=0$ and $2\pi$) and also vanishes on its boundaries.
Hence $\bar{\Lambda}^{(1)} \in {\cal T}^{(1)}$,
${\cal G}(\bar{\Lambda}^{(1)})$ annihilates physical states and
$|\frac{1}{2}>$ is gauge invariant.

The standard vertex operator $U(\theta_{0})$ is not quite $V$, but an operator
creating a blip state from $|0>$ localized at some angle $\theta_{0}$.  We can
make up this operator from a function $\bar {\chi}$ which equals $2\pi
\Lambda i\frac {\tau_{3}}{2}$ on $\partial
H$ and $\Theta i \frac {\tau_{3}}{2}$ on $\partial D$, $\Lambda$ being defined
before (3.3) and $\Theta$ after (3.6).
The U(1)
computation of Section 3 gives us its regularized form on suitably identifying
the U(1) generated by $\frac{\tau_{3}}{2}$ with the U(1) of that Section :
\be
U(\q_{0})=:e^{iQ(\bar{\chi})}:~~,~~~
Q(\bar{\chi})=\frac{1}{2\pi}\int {\rm Tr} (-d\bar{\chi}A+\bar{\chi}A^{2})~.
\ee

The KM representation associated with $U(\theta_0)|0>$ is known to be the same
as the one defined by $|\frac {1}{2}>$.

As we remarked in Section 3, the Fubini-Veneziano
``position'' and ``momentum'' operators occur naturally in CS theory.  This is
especially so for $Q(\underline{\theta})$ which is a combination of
position and momentum.

There is a construction, the ``vertex operator construction,'' of all KM
generators from (5.17) \cite{godd}.   As their algebraic properties and
action on
$|\frac{1}{2}>$ are exactly the same as those of
$H^{(1)\alpha}_{N},~E_{N}^{(1)\alpha}$
defined here, it must be so that both are (weakly) identical.  A direct
demonstration of this identity will not be attempted in this paper.

The CS diffeo generators for the conformal family at
$\partial H$ (constructed along the lines of ref.
[11]) become, in a standard notation \cite{godd},
the Virasoro generators in quantum theory and are given by the
Sugawara construction. A proof of this result can be developed using our U(1)
treatment \cite{bim}. We will hereafter write $L^{(1)}_N$ for these $L_N$ to
indicate the fact that they correspond to zero vector fields on $\partial D$.

Let $\tilde {U}$ be the operator creating blips at both $z \in \partial H$
 and $P \in \partial D$.
The work in\\
Section 3 shows how to construct it by using the function $2\pi\Lambda i
\frac{\tau_3}{2}$ instead of $\bar{\chi}$.
This operator appears to be related to the regularized version of the path
ordered integral
\be
W_{\frac{1}{2}\frac {1}{2}}=
[P\exp \left(-\int A \right )]_{\frac {1}{2}\frac {1}{2}}
\ee
where $A=iA^{\alpha}_{i} \frac{1}{2}\tau_{\alpha}dx^{i}$, the
integration is along a line $L$ defined following Section 3 and
the matrix elements of the Wilson matrix
$$ W=P\exp \left(- \int A \right )$$
are between $\frac{\tau_{3}}{2}$
eigenstates for eigenvalue $\frac {1}{2}$.  This conjecture is made plausible
by the following: i) The response of a state to $W$ is insensitive to
deformations of $L$ (keeping $z,~P$ and
tangents there fixed), a property it shares
with $\tilde {U}$. ii) The action of a gauge transformation $g$ on
$W$ is $W\rightarrow g(z) W g^{-1}(P)$.
$W$ thus commutes with ${\cal G} (\Lambda^{(1)})$,
another property it shares with $\tilde {U}$.
iii) This gauge response of $W$ also shows that $W_{\frac {1}{2} \frac
{1}{2}}|0>$ is the highest weight state for $H^{(1)3}_{0}=\frac{1}{2}$ of the
global SU(2) localized on $\partial H$.  So obviously
is $\tilde {U}|0>$.
iv) We have already seen in Section 3 that there is a
precise correspondence between the Wilson integral and $\tilde {U}$
for the group U(1).

The vertex operator construction for all KM algebras based on compact simple
groups is known from the point of view of the theory of these algebras.  With
this knowledge in mind, it is easy to extrapolate the preceding CS approach to
the SU(2) KM vertex operator theory to these algebras as well.

We conclude this subsection by enquiring about the sense in which the vertex
operator construction describes
a source with internal vector $I$ for a general $G$.
Towards this end, let us first recall how the U(1) CS constraint
$F_{12}(x)=-\frac{2\pi e}{k} \delta^{2}(x-z)$ is
interpreted and enforced in quantum theory.  We
first rewrite it in the form
\be
-\frac{k}{2\pi}\oint_{\cc}A =e
\ee
where $\cal {C}$ is any positively oriented contour enclosing $z$.
The left hand side here is next identified with $q^{(1)}_{0} = q(\xi^{(1)}_0)$
which follows classically from Stokes' theorem and partial integration of
(2.5), since $F_{12}$ is numerically zero on $D \setminus H$.
Thus, (5.19) is $q^{(1)}_{0}= e$.  As $q^{(1)}_{0}$ is
the classical charge and the quantum charge is $Q^{(1)}_{0}$, the quantum
version of (5.19) is
taken to mean that $Q^{(1)}_{0}=e$ on all states.  We finally verify
that this equality is fulfilled on the states obtained by the vertex
operator approach.

Somewhat similar tactics can be pursued for nonabelian CS theories.  For this
purpose, we first rewrite (5.3) in an integral form using the nonabelian Stokes
theorem [see ref. 3 which also contains citations to the original work on this
theorem]:
\be
P\exp\left (-\int_{\cc(x_{0})}A \right )
=U^{-1}_{{\cal L}}[\exp ( i \frac{2\pi}{k}I)]~
U_{{\cal L}}
\ee
${\cal C}(x_{0})$ here is a positively oriented
contour around $z$ starting and terminating at
$x_{0}$, ${\cal L}$ is a line from $x_{0}$ to $z$ and
\be
U_{{\cal L}}=P\exp \left (-\int_{{\cal L}} A \right ).
\ee

Now the left integral in (5.20)
does not change if ${\cal L}$ approaches $z$
in different directions.  Hence the change of $U_{{\cal L}}$ as the
angle of approach to $z$ is
changed is of the form $U_{{\cal L}}\rightarrow H U_{{\cal L}}$
where $H$ is in the
stability group of $I:H^{-1}IH=I$ \cite{bal}.  Therefore $A_{\theta}$ on
$\partial H$  is in the Lie algebra of the stability group of $I$ \cite{bal}.

It is known that all UIR's of $G$ can be obtained from $I$ with stability
groups
generated by Cartan subalgebras.  We therefore assume that the stability group
$T$ of $I$ has a Cartan subalgebra $\underline C$ as its Lie algebra.
$A_{\theta}$ is then $\underline C$ valued on $\partial H$.

Now consider the limit where ${\cal C} (x_{0})$ shrinks to $\partial H$.
$U_{{\cal L}}$
tends to an element of $T$ in that limit and drops out of (5.20) while the path
ordering there also becomes redundant.  Taking logarithms, we can now rewrite
it as
\be
\oint_{\partial H} A = -i \frac{2\pi}{k} I\;.
\ee
Another way to get (5.22) is to first restrict (5.3) to $\partial H$.  This
alternative also justifies our taking logarithms.

Following our abelian approach, we now declare the quantum version of (5.3) to
be the equality
\be
Q^{(1)\alpha}_{0}=\hat{I}_{\alpha}~~,
\ee
$\hat {I}_{\alpha}$ being the quantum operators for the classical $I_{\alpha}=
{\rm Tr}~T_{\alpha}I$.
They fulfill the commutation relations appropriate to $\uG$ and generate the
UIR associated with $I$ \cite{marmo}.
For SU(2), if the orbit of $I$ contains $j
\tau_{3}$, then $\hat{I}_{\alpha}$ generate the spin $j$ representation.

Let us now specialize to SU(2), considerations for general $G$ being similar.
Let us also initially limit ourselves to the orbit of $\frac{1}{2}\tau_{3}$ so
that $j=\frac {1}{2}$.  We have seen that the  SU(2) UIR for a highest weight
KM state has $j=\frac {1}{2}$.  On this family of states, we have then an
enforcement of the relation (5.23).  Repeated application of KM generators on
states describing the $\frac {1}{2}$ UIR also generates any $j$ of the form
$\frac {1}{2} + K, K \in \bf Z^{+}$, so that the relation (5.23) for any $j$ is
enforced on an appropriate subset of states.

The KM states describe all $j=\frac {1}{2},~\frac {3}{2},\ldots$ so that (5.23)
with a fixed $j$ can not be enforced on all states.  This large family of
states makes its appearance because of our regularization of the Gauss law
(5.3).  We may of course discard all states except those with a fixed $j$, but
at the cost of losing diffeomorphism invariance on $\partial H$.  This is
because a diffeo generates $Q^{(1)\alpha}_{N}$ from $Q^{(1)\alpha}_{0}$,
the commutator $[L^{(1)}_{N},Q_{0}^{(1)\alpha}]$ of
the Virasoro generator $L^{(1)}_N$ with $Q^{(1)\alpha}_{0}$ being proportional
to $Q^{(1)\alpha}_{N}$. [ The superscript $1$ on $L^{(1)}_N$ is to indicate
that it corresponds to the zero vector field on $\partial D$.]

The state $V|0>$ is an eigenstate of $L^{(1)}_0$. The corresponding eigenvalue
$\frac{1}{2+4}\frac{3}{2}=\frac{1}{4}$ follows from the formula \cite{godd}
\be
L^{(1)}_{0}=\frac{1}{2k+c_{V}}\sum_{\alpha}
\sum_{N}:Q^{(1)\alpha}_{-N}Q^{(1)\alpha}_{N}:~~,
\ee
$c_{\rm {v}}$
being the quadratic Casimir operator in the adjoint representation.
The operator $L^{(1)}_{0}$ generates rotation diffeos of $\partial H$. $V|0>$
thus has spin $S=\frac {1}{4}$.
It is also associated with the $\frac {1}{2}$ representation of
$G=SU(2)$. Note that in
the KM representation space, there will in general be multiple
occurrences of a given UIR of $G$, possibly differing in spin.
Note also that there is a
spin $S-$ internal symmetry $j$ correlation predicted by a KM UIR which may
have
some phenomenological use.

States compatible with (5.23) for
integer $j$ do not require a vertex operator for construction.  They
can be obtained by applying polynomials in $Q^{(1)\alpha}_{N}$ to $|0>$.
Remarks similar to the preceding ones can be made about these states as well.

\newpage
{\bf 5.2.  The Spin-Statistics Theorem}

We restrict ourselves to level 1 representations and for clarity consider
$G=SU(2)$.  The proof is accomplished using the earlier abelian ideas, so we
can be brief.

We first examine a two particle state consisting of two identical blips at
$z^{(i)}~(i=1,2)$ in $\tau_3$ direction (and correspondingly, two blips at
$z^{(i)^\prime}$).  It is obtained by applying vertex operators
$\tilde{U}$ to the tensor product of highest weight states with zero
$Q_{0}^{(i)\alpha}$ and corresponds to a state in the tensor product of level 1
representations with $j=\frac{1}{2}$ highest weight states.
Figure 2 is a visual display of this state.  When the
particles are exchanged without disturbing the internal states,
 we get Figure 4b
which as before is equal to Figures 4c and 4d. The canonical spin-statistics
connection is thus established for these states.

The general two particle state with identical internal states
in this tensor product space is obtained from
a preceding state by applying two identical polynomials ${\cal P}$
of the KM generators
$Q^{(i)\alpha}_{N}$ for the holes at $z^{(i)}$ [and perhaps also taking limits
of such states].  The operators ${\cal P}(\{Q^{(1)\alpha}_{N}\})$ and
${\cal P}(\{Q^{(2)\alpha}_{N}\})$ are commuting tensorial local fields.  The
spin-statistics connection is thus valid for these more general states.

This remark also shows the validity of this theorem for integer $j$ states
obtained from the tensor product of
the highest weight KM states $|0>_{(i)}$ localized
at $z^{(i)}$ (and with zero $Q^{(i)\alpha}_{0})$ by applying
${\cal P}(\{Q^{(1)\alpha}_{N}\})$ ${\cal P}(\{Q^{(2)\alpha}_{N}\})$.
The exchange operator
$\sigma$ is clearly trivial here.  Further, the $2\pi$ rotation for hole 1 say
is $\exp (i 2\pi L^{(1)}_{0})$
where the  Virasoro generator $L^{(1)}_{0}$ is the
spin operator for hole 1.  It commutes with the polynomials and is one on
$|0>_1 \bigotimes |0>_2$. The theorem
$ \sigma = \exp (i 2\pi L^{(1)}_0)$ thus follows.

Suppose next that the internal states of the particles differ and that the
two-body state is
$A {\cal P}(Q^{(1)\alpha}_N) {\cal P}(Q^{(2)\alpha}_N) |0>_1 \bigotimes
|0>_2$.  $A$ here is an element of the KM group for hole 1, and hence commutes
with $Q^{(2)\alpha}_{N}$ and is 1 on $|0>_{2}$.  This state is similar to
$|~{\rm Proton}>~~|~ {\rm Neutron}>$ in
a nucleon model with isospin symmetry, the role of isospin
SU(2) being assumed here by the KM group.  The exchange operator for this state
is $A \sigma A^{-1}=A \exp (i2\pi L^{(1)}_{0})A^{-1}$. But $\exp (i 2\pi
L^{(1)}_{0})$ commutes with $A$ and  hence so does $\sigma$.  Therefore, the
$2\pi$  rotations and exchanges are equal to their old versions and mutually
equal.  A similar result is readily proved for $A$ times the state involving
the vertex operators looked at previously.

A completely general state is a linear
combination of states of the sort considered in the
preceeding paragraphs. We thus
have the standard spin-statistics connection for level 0 and 1 KM states.  But
we do not have a proof for a general KM representation.

There seems to be no particular difficulty in extending this discussion to CS
theories for all compact simple groups.  The details will be omitted here.

{\bf Acknowledgements}

We have been supported during the course of this work as follows:~1) A.~P.~B.,
G.~B.,~K.~S.~G.~ by the Department of Energy, USA, under contract number
DE-FG-02-85ER40231,
and A.~S.~by the Department of Energy, USA under contract
number DE-FG05-84ER40141;
2) A.~P.~B.~and A.~S.~ by INFN, Italy [at Dipartimento di
Scienze Fisiche, Universit{\`a} di Napoli]; 3) G.~B.~ by the Dipartimento
di Scienze Fisiche, Universit{\`a} di Napoli.  We wish to thank the
 group in Naples and Giuseppe Marmo, in particular, for their hospitality while
this work was in progress. We also wish to thank Paulo Teotonio for very
helpful
comments.

\end{document}